\documentclass{article}[9pt]
\usepackage{amsmath}
\usepackage{amssymb}
\usepackage{amsfonts}
\usepackage{graphicx}
\usepackage{enumerate,color,graphicx,fancybox,pifont,epsf,epsfig,subfigure,amsmath,amssymb,psfrag,spconf}
\usepackage{algorithm}
\usepackage{algorithmic}
\usepackage{cite}
\newcounter{MYtempeqncnt}

\newcommand{\figref}[1]{Figure~\ref{#1}}

\newcommand{\algref}[1]{Algorithm~\ref{#1}}

\newcommand{\qed}{\nobreak \ifvmode \relax \else
      \ifdim\lastskip<1.5em \hskip-\lastskip
      \hskip1.5em plus0em minus0.5em \fi \nobreak
      \vrule height0.75em width0.5em depth0.25em\fi}

\title{ANALYSIS-BY-SYNTHESIS-BASED QUANTIZATION OF \\ COMPRESSED SENSING MEASUREMENTS}
\name{Amirpasha Shirazinia, Saikat Chatterjee, Mikael Skoglund}
\address{Communication Theory Lab, ACCESS Linnaeus Centre \\
KTH Royal Institute of Technology, Stockholm, Sweden\\
{\small \tt Email: amishi@kth.se, sach@kth.se, skoglund@kth.se}}

\begin{document}
\maketitle
\ninept
\begin{abstract}
We consider a resource-constrained scenario where a compressed sensing- (CS) based sensor has a low number of measurements which are quantized at a low rate followed by transmission or storage. Applying this scenario, we develop a new quantizer design which aims to attain a high-quality reconstruction performance of a sparse source signal based on analysis-by-synthesis framework. Through simulations, we compare the performance of the proposed quantization algorithm vis-a-vis existing quantization methods.
\end{abstract}


\vspace{-0.3cm}
\section{Introduction} \label{sec:intro}
\vspace{-0.1cm}
Using a model of under-determined linear set of equations, compressed sensing (CS) \cite{08:Candes} aims to reconstruct a high-dimensional sparse source vector (where most of coefficients are zero) from an under-sampled low-dimensional measurement vector. With a limited number of measurements (or a limited resource of sampling), CS has emerged as a new powerful tool for sparse signal acquisition, compression and reconstruction. In many practical applications, CS measurements need to be quantized into a finite resolution representation, and then transmitted to a destination point for sparse signal reconstruction followed by other inference tasks. In this paper, we consider application scenarios where both measurement (or sampling) and transmission resources are constrained. For transmission resource, we mean that the available bits to quantize the CS measurements are limited. Considering availability of limited number of measurements and quantization bits, we design new quantization algorithms where our goal is to achieve high quality sparse signal reconstruction from the quantized CS measurements.

CS with quantized measurements has recently started to gain significant attention in literature, and most commonly, the focus in this area is on three main categories: (1) extensions to existing CS reconstruction algorithms while quantization schemes remain unchanged \cite{06:Candes2,10:Sinan,10:Zymnis,11:Dai,11:Jacques,12:Yan,12:Kamilov}. (2) In the second category, trade-offs between the aspects of quantization (e.g., quantization rate) and CS (e.g., number of measurements and loss in sparse reconstruction) have been considered \cite{08:Goyal,11:Dai,12:Laska}. (3) In another important class, the main concentration is on quantizer design for CS measurements while CS reconstruction methods are fixed \cite{09:Sun,11:Kamilov,12:Boufounos,13:Pasha2,12:Pasha1}.

The main contribution of this work is in the third category mentioned above, i.e., quantizer design for CS measurements while CS reconstruction methods are fixed. We develop a new framework for scalar quantization of CS measurements with the objective of achieving a lower \textit{end-to-end reconstruction distortion} rather than \textit{quantization distortion} for CS measurements. Technically, given a fixed quantizer look-up table and a CS reconstruction algorithm, our proposed algorithm strategically employs a two-stage mechanism in a closed-loop: (1) the synthesis step uses a sparse signal reconstruction technique for measuring the direct effect of quantization of CS measurements on the final sparse signal reconstruction quality, and (2) the analysis step decides appropriate quantized values to maximize the final sparse signal reconstruction quality. This closed-loop strategy is known as \textit{analysis-by-synthesis} (AbS) which has been widely used in multi-media coding \cite{88:Kroon,95:Aizawa,97:George}. To the best of our knowledge, the AbS approach has not been used for quntization of CS measurements, where we show by exploiting this framework, a significantly better reconstruction performance is provided compared to the schemes which only consider quantization distortion, but at the expense of a higher computation. We analyze computational complexity of the proposed algorithm, where it is shown that the complexity depends upon the availability of two compression resources, i.e.,  quantization bit rate, and number of CS measurements.

\textit{Notations:} Scalar random variables (RV's) will be denoted by upper-case letters and their instants by the respective lower-case letters. Random vectors will be represented by boldface characters. Further, a set is shown by a calligraphic character and its cardinality by $|\cdot|$. We will also denote the transpose of a vector by by $(\cdot)^T$. We will use~$\mathbb{E}[\cdot]$ to denote the expectation operator. The $\ell_p$-norm ($p \geq 0$) of a vector will be denoted by $\|\cdot\|_p$. 
\vspace{-0.3cm}
\section{Problem Statement} \label{sec:problem}
\vspace{-0.1cm}

\subsection{Preliminaries of CS Framework} \label{subsec:formulation}
\vspace{-0.2cm}
Formally, we let a random sparse (in a fixed basis) signal $\mathbf{X} \in \mathbb{R}^M$ be linearly encoded using a known deterministic sensing matrix $\mathbf{\Phi} \in \mathbb{R}^{N \times M}$ ($N < M$) representing measuring (sampling) system which results in an under-determined set of linear measurements $\mathbf{Y = \Phi X} \in \mathbb{R}^N$. We let $\mathbf{X}$ be a $K$-sparse vector, i.e., it has at most $K$ ($K < N$) non-zero coefficients, where the location and magnitude of the non-zero coefficients are drawn randomly from known distributions. We also note that the sparsity level $K$ is known in advance. We define the support set of the sparse vector $\mathbf{X} = [X_1,\ldots,X_M]^T$ by $\mathcal{S} \triangleq \{m : X_m \neq 0 \} \subset \{1,\ldots,M\}$ with $|\mathcal{S}| = \|\mathbf{X}\|_0 \leq K$.

In order to estimate a sparse source vector from under-determined linear measurements, several efficient techniques have been developed based on convex optimization (see e.g.~\cite{06:Candes2,07:Candes}), iterative greedy search (see e.g. \cite{07:Tropp,08:Blumensath,09:Dai,12:Saikat}) and Bayesian estimation approaches (see e.g. \cite{07:Larsson,08:Ji,09:Elad,10:Protter}). The results of this paper are generic, and we do not use a specific CS reconstruction algorithm. Denoting a sparse reconstruction function by $\textsf{R}$, it is defined by a mapping $\textsf{R}: \mathbb{R}^N \rightarrow \mathbb{R}^M$ which takes a (possibly corrupted) measurement vector in $N$-dimensional space, and produces an estimate of the sparse source vector in $M$-dimensional space ($N<M$). 
\vspace{-0.25cm}
\subsection{Quantization of CS Measurements} \label{subsec:SQ}
\vspace{-0.1cm}
We consider scalar quantization of the random CS measurements $Y_n$'s ($n=1,\ldots,N$). For this purpose, quantization is divided into \textit{encoding} and \textit{decoding} tasks. We consider a scalar \textit{quantizer encoder} which maps each measurement to an appropriate index in a finite integer set in order for a \textit{quantizer decoder} to make an estimate of the measurements based on the received index and a known decoding look-up table. We assume that the total bit budget (rate) allocated for quantization is $R_x \triangleq M r_x$ bits per vector $\mathbf{X}$ in which $r_x \in \mathbb{R}^+$ is the assigned quantization rate to a scalar component of $\mathbf{X}$. Having the observations $\mathbf{Y = \Phi X}$, each entry of the measurement vector, $Y_n$ ($n=1,\ldots,N$), is encoded via $r_y \triangleq Mr_x/N$ bits. For each entry $Y_n$, a quantizer encoder is defined by a mapping $\textsf{E}: \mathbb{R} \rightarrow \mathcal{I}$, where $\mathcal{I}$ denotes the index set defined as $\mathcal{I} \triangleq \{0,1,\ldots,2^{r_y}-1\}$ with $|\mathcal{I}| = 2^{r_y}$. Denoting the quantized index by the RV $I_n$ ($n =1,\ldots,N$), the encoder acts according to $Y_n \in \mathcal{R}^{i_n} \Rightarrow I_n = i_n$, where the sets $\{\mathcal{R}^{i_n}\}_{i_n=0}^{2^{r_y}-1}$ are called encoder regions and $\bigcup_{i_n=0}^{2^{r_y}-1} \mathcal{R}^{i_n} = \mathbb{R}$. In words, when $Y_n$ belongs to the region $\mathcal{R}^{i_n}$, the encoder picks the index $i_n \in \mathcal{I}$. Next, we define quantizer decoder which is characterized by a mapping $\textsf{D}: \mathcal{I} \rightarrow \mathcal{C}_n$. The quantizer decoder takes the index $I_n$, and performs according to an available look-up table; $I_n \!=\! i_n \Rightarrow \widehat{Y}_n \!=\! c_{i_n}$ such that when the received index is $i_n$, the decoder outputs the codepoint $c_{i_n}$. Note that $\widehat{Y}_n$ is the quantized measurement RV associated with the entry $Y_n$, and the set of all reproduction \textit{codepoints} $\mathcal{C}_n \! \triangleq \! \{c_{i_n}\}_{i_n\!=\!0}^{2^{r_y}-1}$ associated with this entry is called a \textit{codebook}. We denote by $\widehat{\mathbf{X}} \!=\! \textsf{R}\left([c_{I_1}\!,\!\ldots\!,\!c_{I_N}]^T\right) \!\in\! \mathbb{R}^M $ the estimation of the source from the quantized measurements using a CS reconstruction function \textsf{R}.
\vspace{-0.25cm}
%%%%%%%%%%%%%%%%%%%%%%%%%%%%%%%%%%%%%%%%%%%%%%%%%%%%%%%%%%%%%%%%%%%%%%%%%%%%%%%%%
\subsection{Objective and Performance Criterion} \label{subsec:criteria}
%%%%%%%%%%%%%%%%%%%%%%%%%%%%%%%%%%%%%%%%%%%%%%%%%%%%%%%%%%%%%%%%%%%%%%%%%%%%%%%%%
\vspace{-0.1cm}
In this paper, we are interested in addressing the following quantizer design problem: Given a CS measurement vector $\mathbf{Y = \Phi X} \in \mathbb{R}^N$, a CS reconstruction function $\textsf{R}$ and codebook sets $\mathcal{C}_n = \{c_{i_n}\}_{i_n=0}^{2^{r_y}-1}$ ($n=1,\ldots,N$) for a fixed bit budget $R_x$, the objective is to find encoding indexes $i_n \in \mathcal{I}$ ($n=1,\ldots,N$), such that the end-to-end MSE of the estimated vector $\widehat{\mathbf{X}} \in \mathbb{R}^M$, i.e.  $\mathbb{E}[\|\mathbf{X - \widehat{X}}\|_2^2]$, is minimum. In other words, we address the optimization problem
\begin{equation} \label{eq:opt e2e}
    \{i_1^\star,\ldots,i_N^\star\} = \underset{\{i_n \in \mathcal{I}\}_{n=1}^N}{\textrm{argmin }} \mathbb{E} [\|\mathbf{X} - \widehat{\mathbf{X}} \|_2^2 ],
    \vspace{-0.15cm}
\end{equation}
where $\{i_n^\star\}_{n=1}^N$ are the optimal encoding indexes (w.r.t. to minimizing the end-to-end MSE  given codebook sets) for quantization of the measurement vector $\mathbf{Y}=[Y_1,\ldots,Y_N]^T$. Also, note that the end-to-end distortion $\mathbb{E} [\|\mathbf{X} - \widehat{\mathbf{X}} \|_2^2 ]$ depends upon CS reconstruction distortion as well as quantization distortion.

In this paper, our aim is higher than just minimizing the \textit{quantization distortion} $\mathbb{E}[\|\mathbf{Y} - \widehat{\mathbf{Y}}\|_2^2]$ considered in the design of \textit{nearest-neighbor coding} where each measurement entry is coded to its nearest codepoint. The nearest-neighbor coding does not necessarily guarantee that the \textit{end-to-end distortion}, i.e., $\mathbb{E} [\|\mathbf{X} - \widehat{\mathbf{X}} \|_2^2]$, is also minimized subject to fixed codebook sets. This is due to nonlinear behavior of CS reconstruction algorithms and non-orthogonality of the CS system.

Unfortunately, solving \eqref{eq:opt e2e} jointly for all encoding indexes is not analytically and practically feasible for a generic sparse reconstruction algorithm since it is performed by searching over all possible $2^{Mr_x}$ codepoints, leading to high complexity. Instead, in this work, we focus on a suboptimal technique for quantization of CS measurements which is computationally efficient and also provides a high-quality reconstruction performance. 
\vspace{-0.35cm}
%%%%%%%%%%%%%%%%%%%%%%%%%%%%%%%%%%%%%%%%%%%%%%%%%%%%%%%%%%%%%%%%%%%%%%%%%%%%%%%%%%%%
\section{Analysis-by-Synthesis Quantization of CS Measurements} \label{sec:quant}
\begin{figure}
\begin{center}
  \psfrag{x}[][][0.65]{$\mathbf{X}$}
  \psfrag{Psi}[][][0.8]{$\mathbf{\Phi}$}
  \psfrag{y}[][][0.6]{$\mathbf{Y}$}
  \psfrag{E}[][][0.7]{\hspace{0.25cm} AbS quantizer encoder}
  \psfrag{CS_E}[][][0.7]{CS encoder}
  \psfrag{D}[][][0.7]{\hspace{-0.2cm} Quantizer decoder}
  \psfrag{Q}[][][0.75]{$\textsf{E}$}
  \psfrag{C_1}[][][0.65]{$\mathcal{C}_1$}
  \psfrag{C_N}[][][0.65]{$\mathcal{C}_N$}
  \psfrag{I_1}[][][0.6]{$I_1$}
  \psfrag{I_N}[][][0.6]{$I_N$}
  \psfrag{y_1}[][][0.6]{$Y_1$}
  \psfrag{y_N}[][][0.6]{$Y_N$}
  \psfrag{Q-1}[][][0.75]{$\textsf{D}$}
  \psfrag{y_h_1}[][][0.6]{$\widehat{Y}_1$}
  \psfrag{y_h_N}[][][0.6]{$\widehat{Y}_N$}
  \psfrag{y_h}[][][0.6]{$\widehat{\mathbf{Y}}$}
  \psfrag{Rec}[][][0.65]{CS decoder}
  \psfrag{R}[][][0.8]{\textsf{R}}
  \psfrag{x_h}[][][0.65]{$\widehat{\mathbf{X}}$}
  \includegraphics[width=9cm, height=4cm]{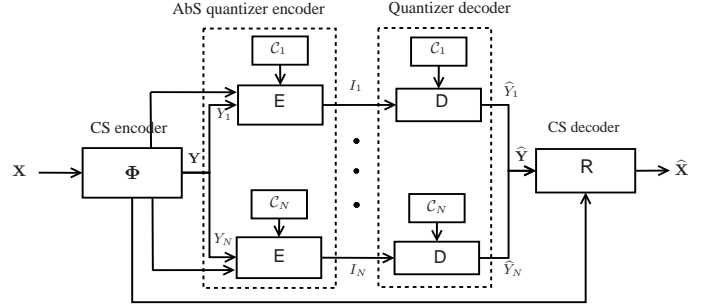}\\
  \vspace{-0.35cm}
  \caption{AbS quantization of CS measurements.}
  \label{fig:diagram_CS}
  \end{center}
  \vspace{-0.5cm}
\end{figure}

%%%%%%%%%%%%%%%%%%%%%%%%%%%%%%%%%%%%%%%%%%%%%%%%%%%%%%%%%%%%%%%%%%%%%%%%%%%%%%%%%%%%
\vspace{-0.1cm}
Our proposed AbS-based quantization system is illustrated in \figref{fig:diagram_CS}.
In order to feasibly solve \eqref{eq:opt e2e}, we
consider optimizing one variable by fixing the others, that is, optimizing the index $i_n$ by fixing the indexes $i_1,\ldots,i_{n-1},i_{n+1},\ldots,i_{N}$ using an \textit{alternating optimization} approach which is potentially suboptimal compared to a joint optimization method, but provides a feasible solution. In what follows, we show how an MSE-minimizing transmission index can be chosen by fixing the others. 
\vspace{-0.3cm}
%%%%%%%%%%%%%%%%%%%%%%%%%%%%%%%%%%%%%%%%%%%%%%%%%%%%%%%%%%%%%%%%%%%%%%%%%%%%%%%%%%%%
\subsection{Optimizing Encoding Indexes} \label{subsec:optimzied índex}
%%%%%%%%%%%%%%%%%%%%%%%%%%%%%%%%%%%%%%%%%%%%%%%%%%%%%%%%%%%%%%%%%%%%%%%%%%%%%%%%%%%%
\vspace{-0.2cm}
Let us first rewrite the end-to-end MSE, $\mathbb{E}[\|\mathbf{X} - \widehat{\mathbf{X}}\|_2^2]$, as \eqref{eq:suboptimal condition}, where $(a)$ is followed by marginalization over $I_n$ and $\mathbf{Y}$. Also, $(b)$ follows from interchanging the integral and summation and the fact that $\textrm{Pr} \{I_n=i_n | \mathbf{Y=y}\} = 1$, $\forall y_n \in \mathcal{R}^{i_n}$, and otherwise the probability is zero. Moreover, $f(\mathbf{y})$ is the $N$-fold probability density function (p.d.f.) of the measurement vector. Note that by denoting $\widehat{\mathbf{X}}(I_n) \triangleq \textsf{R}\left([c_{i_1},\ldots,c_{I_n},\ldots,c_{i_N}]\right)^T$, we imply that the reconstructed signal is dependent only upon the index (equivalently codepoint) associated with the $n^{th}$ measurement entry. Now, let denote the \textit{minimum mean square error} (MMSE) estimation of $\mathbf{X}$ given the measurements $\mathbf{Y}=\mathbf{y}$ by
\begin{equation} \label{eq:MMSE est}
    \tilde{\mathbf{x}}(\mathbf{y}) \triangleq \mathbb{E}[\mathbf{X|Y=y}] \in \mathbb{R}^M,
\end{equation}
then, given the fixed codebooks $\mathcal{C}_n$ $(n=1,\ldots,N)$, the MSE-minimizing index (assuming other indexes are fixed) is identical to finding the index that minimizes the term in the braces in the last expression of \eqref{eq:suboptimal condition} since $f(y)$ is non-negative. The resulting index denoted by $i_n^\star \in \mathcal{I}$ is given by  \eqref{eq:proof enc}, where $(a)$ follows from the fact that $\mathbf{X}$ is independent of $I_n$ conditioned on $\mathbf{Y}$, hence, $\mathbb{E} \left[\|\mathbf{X}\|_2^2 | \mathbf{Y\!=\!y}, I_n\!=\!i_n \right] = \mathbb{E} \left[\|\mathbf{X}\|_2^2 | \mathbf{Y\!=\!y}\right]$ which is pulled out of the optimization. Also, $(b)$ follows from a similar rationale, i.e., $\widehat{\mathbf{X}}(I_n)$ is independent of $\mathbf{Y}$ given $I_n$. Further, $\mathbf{X}$ and $\widehat{\mathbf{X}}(I_n)$ are independent given $\mathbf{Y}$ and $I_n$.

\begin{figure*}[!t]
\normalsize
\setcounter{MYtempeqncnt}{\value{equation}}
\begin{equation} \label{eq:suboptimal condition}
\begin{aligned}
    \mathbb{E}[\|\mathbf{X} - \widehat{\mathbf{X}} \|_2^2] &= \mathbb{E}[\|\mathbf{X} - \widehat{\mathbf{X}}(I_n) \|_2^2] \stackrel{(a)}{=} \int_{\mathbf{y}} \sum_{i_n} \textrm{Pr}\{I_n = i_n | \mathbf{Y=y}\} \mathbb{E}[\|\mathbf{X} - \widehat{\mathbf{X}}(I_n) \|_2^2 | \mathbf{Y=y},I_n=i_n] f(\mathbf{y}) d \mathbf{y}& \\
    &\stackrel{(b)}{=}  \int_{y_1} \ldots \int_{y_{n-1}} \int_{y_{n+1}} \ldots \int_{y_{N}} \sum_{i_n} \int_{y_n \! \in \! \mathcal{R}^{i_n}} \left\{\mathbb{E}[\|\mathbf{X} - \widehat{\mathbf{X}}(I_n) \|_2^2 | \mathbf{Y=y},I_n=i_n] \right\}f(\mathbf{y}) d \mathbf{y}&
\end{aligned}
\end{equation}
\setcounter{equation}{\value{MYtempeqncnt}}
\hrulefill
\end{figure*}
\setcounter{equation}{3}

\begin{figure*}[!t]
\normalsize
\setcounter{MYtempeqncnt}{\value{equation}}
\begin{equation} \label{eq:proof enc}
\begin{aligned}
    i_n^\star &= \textrm{arg}\underset{i_n \in \mathcal{I}}{\textrm{min }} \mathbb{E}[\|\mathbf{X} - \widehat{\mathbf{X}}(I_n) \|_2^2 | \mathbf{Y=y},I_n=i_n]
    \stackrel{(a)}{=} \textrm{arg}\underset{i_n \in \mathcal{I}}{\textrm{min }} \left\{\mathbb{E}[\|\widehat{\mathbf{X}}(I_n)\|_2^2  | \mathbf{Y\!=\!y}, I_n=i_n] \!-\! 2\mathbb{E}[\mathbf{X}^T \widehat{\mathbf{X}}(I_n)  | \mathbf{Y=y}, I_n=i_n] \right\}& \\
    &\stackrel{(b)}{=} \textrm{arg}\underset{i_n \in \mathcal{I}}{\textrm{min }} \left\{\mathbb{E}[\|\widehat{\mathbf{X}}(I_n)\|_2^2 \big | I_n=i_n] - 2\mathbb{E}[\mathbf{X}^T \big | \mathbf{Y=y}] \mathbb{E}[\widehat{\mathbf{X}}(I_n) \big | I_n=i_n] \right\}&
\end{aligned}
\end{equation}
\setcounter{equation}{\value{MYtempeqncnt}}
\hrulefill
\end{figure*}
\setcounter{equation}{4}
Following \eqref{eq:MMSE est}, the last expression in \eqref{eq:proof enc} can be rewritten as
\begin{equation} \label{eq:final enc}
\begin{aligned}
    i_n^\star &= \textrm{arg}\underset{i_n \in \mathcal{I}}{\textrm{min}} \left\{\|\widehat{\mathbf{x}}(i_n) \|_2^2- 2 \tilde{\mathbf{x}}(\mathbf{y})^{T} \widehat{\mathbf{x}}(i_n) \right\}.&
\end{aligned}
\end{equation}
One method to predict the MSE-minimizing encoding index is to find the codepoint which after passing through a sparse reconstruction algorithm reproduces a signal vector that is the best estimation to the current input signal vector. Interestingly, \eqref{eq:final enc} implies such an \textit{analysis-by-synthesis} (AbS) method. We use this principle to first find the optimized encoding index for each measurement entry separately (while others are fixed given the codepoints), and then combine them in an alternate-iterate procedure which will be described in details in the next section. Note that we also assume that the codebook sets are available at the encoder as well as the decoder.

\vspace{-0.3cm}
%%%%%%%%%%%%%%%%%%%%%%%%%%%%%%%%%%%%%%%%%%%%%%%%%%%%%%%%%%%%%%%%%%%%%%%
\subsection{Proposed Quantization Algorithm} \label{subsec:algorithm}
%%%%%%%%%%%%%%%%%%%%%%%%%%%%%%%%%%%%%%%%%%%%%%%%%%%%%%%%%%%%%%%%%%%%%%%
\vspace{-0.15cm}
We first describe the framework of the proposed quantization method summarized in \algref{alg1}. Suppose that the codebook sets $\mathcal{C}_n$ ($n=1,\ldots,N$) are designed offline, and let the quantizer encoder have access to the sensing matrix $\mathbf{\Phi}$ and sparsity level $K$ as well as the measurements $\mathbf{y}$ (step (1)). In our formulations (e.g. \eqref{eq:final enc}), the MMSE estimator is required, however, in practice, implementing the MMSE estimator may not be feasible. Therefore, in order to obtain a locally reconstructed vector $\tilde{\mathbf{x}}(\mathbf{y})$, we will approximate the MMSE estimator by the output of the low-complexity greedy \textit{orthogonal matching pursuit} (OMP) \cite{07:Tropp,08:Blumensath} reconstruction algorithm (step (2)). Now, we define a dummy vector $\mathbf{z} \in \mathbb{R}^N$ where at the first iteration, its $n^{th}$ component is chosen uniformly at random from the set $\mathcal{C}_{n}$ ($\forall n$) (step (3)). Indeed, the vector $\mathbf{z}$ implies the predicted quantized measurement which is synthesized at the encoder. Throughout iterations, the entries of the vector $\mathbf{z}$ are adjusted towards the directions of the codepoints (in a sequential manner) which give the minimum reconstruction MSE when retrieved using a CS reconstruction algorithm. Now, we describe the subroutine $\texttt{AbS\_seq}(\cdot)$ executed in \algref{alg1}. 
\vspace{-0.2cm}
\begin{algorithm}
\caption{: AbS-based Quantization}\label{alg1}
\begin{algorithmic}[1]
\STATE{\textbf{input: } $\mathcal{C}_n = \{c_{i_n}\}_{i_n=0}^{2^{r_y}-1}$ ($\forall n=1,\ldots,N$) and $\mathbf{\Phi , y}$, $K$, $\gamma$ (stopping threshold)}
\STATE{\textbf{compute: } $\mathbf{\tilde{x}}(\mathbf{y})$ in \eqref{eq:MMSE est}}
\STATE{\textbf{initialize } $\mathbf{z}^{(0)} \in \mathbb{R}^N$, where $z_n^{(0)} \in \mathcal{C}_{n},$ $\forall n$.}
\STATE{Set $l \gets 0$ (iteration counter)}
\REPEAT
    \STATE{$[i_n^\star , \widehat{\mathbf{x}}^{(l+1)}{(i_n^\star)},\mathbf{z}^{(l+1)}] = \texttt{AbS\_seq}(\mathcal{C}_n,\tilde{\mathbf{x}}(\mathbf{y}),\mathbf{z}^{(l)})$ , $\forall n$}
    \STATE{$l \gets l+1$}
\UNTIL{$\left| \|\widehat{\mathbf{x}}^{(l)}(i_n^\star) \|_2^2- 2 \tilde{\mathbf{x}}(\mathbf{y})^{T} \widehat{\mathbf{x}}^{(l)}(i_n^\star)  \right.$ \\
$\hspace{1cm}\left.-\|\widehat{\mathbf{x}}^{(l-1)}(i_n^\star) \|_2^2 + 2 \tilde{\mathbf{x}}(\mathbf{y})^{T} \widehat{\mathbf{x}}^{(l-1)}(i_n^\star) \right| >  \gamma$ , $\forall n$}
%\STATE{$l \gets l-1$}
\STATE{\textbf{output: } $I_n = i_n^{\star}$ , $\widehat{Y}_n = c_{i_n^{\star}}$ , $\forall n$ }
\end{algorithmic}
\end{algorithm}
\vspace{-0.25cm}

\textbf{Sequential AbS quantization} The proposed AbS-based quantization method is summarized in the subroutine $\texttt{AbS\_seq} (\cdot)$ where the main idea is that each measurement entry is sequentially adjusted towards the direction of its MSE-minimizing codepoint at each iteration. Using \algref{alg1}, the function $\texttt{AbS\_seq}(\cdot)$ accepts the codebooks $\mathcal{C}_n$, $\forall n$, the locally reconstructed vector $\tilde{\mathbf{x}}(\mathbf{y})$ and the dummy vector $\mathbf{z}$. At iteration $l$, the $n^{th}$ ($n=1,\ldots,N$) entry of $\mathbf{z}^{(l)}$, denoted by $z_n^{(l)}$, is replaced by all $2^{r_y}$ codepoints from the set $\mathcal{C}_n$ (step (3)) while the other entries are fixed, and the reconstructed vectors, denoted by $\widehat{\mathbf{x}}^{(l)}(i_n) = \textsf{R} (\mathbf{z}^{(l)})$ ($i_n \in \mathcal{I}=\{0,\ldots,2^{r_y}-1\}$), are synthesized corresponding to each vector (step (4)). Then, an optimization is carried out by solving $\underset{i_n \in \mathcal{I}}{\textrm{argmin}} \|\widehat{\mathbf{x}}^{(l)}(i_n)\|_2^2-2\tilde{\mathbf{x}}(\mathbf{y})^{T} \widehat{\mathbf{x}}^{(l)}(i_n)$ so as to find the wining MSE-minimizing encoding index $i_n^\star$ (step (6)). Now, the $n^{th}$ entry of the vector $\mathbf{z}^{(l)}$ is updated by the codepoint associated with the analyzed index, i.e., $c_{i_n^{\star}}$ (step (7)). This procedure continues for each entry of $\mathbf{z}^{(l)}$ sequentially, and the subroutine produces the optimized transmission index $i_n^\star$, and the reconstructed vector $\widehat{\mathbf{x}}^{(l)}(i_n^\star)$ as well as the updated quantized vector $\mathbf{z}^{(l)}$ which will be used by the function at the next iteration of \algref{alg1} (step (9)).

\vspace{0.15cm}
\hrule
\textbf{Subroutine:} $\texttt{AbS\_seq} \left(\mathcal{C}_n , \tilde{\mathbf{x}}(\mathbf{y}),\mathbf{z}^{(l)}\right)$
\vspace{0.15cm}
\hrule
\begin{algorithmic}[1]
    \FOR{$n=1:N$}
        \FOR{$i=0:2^{r_y}-1$}
           \STATE{$z_n^{(l)} \gets c_{i_n}$}
            \STATE{\textbf{compute:} $\widehat{\mathbf{x}}^{(l)}(i_n) = \textsf{R} (\mathbf{z}^{(l)})$}
        \ENDFOR
        \STATE{$i_n^{\star} =  \underset{i_n \in \mathcal{I}}{\textrm{argmin}} \|\widehat{\mathbf{x}}^{(l)}(i_n)\|_2^2-2\tilde{\mathbf{x}}(\mathbf{y})^{T} \widehat{\mathbf{x}}^{(l)}(i_n)$}
        \STATE{\textbf{update:} $z_n^{(l)} \gets c_{i_n^{\star}}(i_n)$}
    \ENDFOR
\STATE{\textbf{output: } $i_n^\star$ , $\widehat{\mathbf{x}}^{(l)}(i_n^\star)$ , $ \mathbf{z}^{(l)}$}
\vspace{0.15cm}
\hrule
\end{algorithmic}
\vspace{0.15cm}

\algref{alg1} iterates until convergence where the stopping criterion is that reconstruction improvement at two consecutive iterations is smaller than a predefined threshold $\gamma > 0$. After convergence, the algorithm outputs the transmission indexes $I_n$'s and the quantized CS measurements $\widehat{Y}_n$'s, $\forall n=1,\ldots,N$, (step (9)). 

Now, we analyze the computational complexity of the proposed quantization method. We quantify how many times a CS reconstruction algorithm is invoked throughout the procedures. First, recall from \eqref{eq:opt e2e} that an exhaustive search for the joint optimization requires $\mathcal{O}(2^{Mr_x})$, or $\mathcal{O}(2^{Nr_y})$ (since $M r_x = N r_y$), computations of a CS reconstruction algorithm which is not permissible in practice. Next, let us consider the sequential AbS-based quantization (Subroutine $\texttt{AbS\_seq}$) at one iteration of \algref{alg1}. The operations for calculating the transmission indexes increase at most like $\mathcal{O}(2^{Mr_x/N} N)$. This implies that for a fixed bit budget $R_x = Mr_x$, by increasing the number of measurements, first the complexity decreases sharply, and then at some point it starts increasing with a small slope. This is due to the fact that the complexity depends on the compression resources, i.e., number of measurements through the linear term $N$ and quantization rate $r_y$ through the exponential term $2^{R_x/N}$.

\vspace{-0.3cm}
%%%%%%%%%%%%%%%%%%%%%%%%%%%%%%%%%%%%%%%%%%%%%%%%%%%%%%%%%%%%%%%%%%%%%%%%%%%%%%%%%%%%%%%
\section{Experiments and Results} \label{sec:numerical}
%%%%%%%%%%%%%%%%%%%%%%%%%%%%%%%%%%%%%%%%%%%%%%%%%%%%%%%%%%%%%%%%%%%%%%%%%%%%%%%%%%%%%%%
\vspace{-0.15cm}

%%%%%%%%%%%%%%%%%%%%%%%%%%%%%%%%%%%%%%%%%%%%%%%%%%%%%%%%%%%%%%%%%%%%%%%%%%%%%%%%%%%%%%%%%%%%
\subsection{Experimental Setups and Results} \label{subsec:setup}
%%%%%%%%%%%%%%%%%%%%%%%%%%%%%%%%%%%%%%%%%%%%%%%%%%%%%%%%%%%%%%%%%%%%%%%%%%%%%%%%%%%%%%%%%%%%
\vspace{-0.1cm}
We quantify the performance using normalized MSE (NMSE) defined as $\textrm{NMSE} \triangleq \frac{\mathbb{E}[\|\mathbf{X}-\widehat{\mathbf{X}}\|_2^2]}{\mathbb{E}[\|\mathbf{X}\|_2^2]}$. In order to measure level of under-sampling, we define the measurement rate as $\alpha \triangleq N/M$ ($0 < \alpha \leq 1$). We choose $\alpha$ for given values of sparsity level $K$ and input vector size $M$, and round the number of measurements $N$ to its nearest integer. We randomly generate a set of $K$-sparse vector $\mathbf{X}$ where the support set $\mathcal{S}$ is chosen uniformly at random over the set $\{1,2,\ldots,M\}$. Non-zero coefficients of $\mathbf{X}$ are drawn according to i.i.d. standard Gaussian random variables. We let the elements of the sensing matrix be $\mathbf{\Phi}_{ij} \overset{\textrm{iid}}{\sim} \mathcal{N} (0,1/N)$, and then normalize the columns of $\mathbf{\Phi}$ to unit-norm. We apply the OMP reconstruction algorithm as a realization of the CS reconstruction function \textsf{R}.

Using simulation parameters $M=512$, $K=35$ (sparsity ratio $\approx 6.8\%$), $r_x = 0.75$ bit per component of $\mathbf{X}$, we have performed 1000 Monte-Carlo simulations to illustrate the performance (NMSE).  For implementation of \algref{alg1} using the subroutine $\texttt{AbS\_seq}$, we choose the stopping threshold $\gamma = 10^{-6}$, where we have observed that \algref{alg1} converges in at most $5$ iterations. We have compared our proposed quantization method (labeled by \textit{``sequential AbS quantization''} in the figures) with relevant methods such as \textit{``nearest-neighbor coding''} of the measurement entries and \textit{``support set coding''} \cite{13:Pasha3}. Employing the support set coding, the $K$ largest non-zero components of $\tilde{\mathbf{x}}$ (in estimated support set) can be represented by $K \log_2 M$ bits, and then their magnitudes are coded to nearest codepoints using $R_x - K \log_2 M$ bits. Another possible scheme is to quantize each component of the reconstructed signal directly using available bit budget. However, our simulations (not included here) have shown that this coding scheme provides a very poor at the decoder, and the remaining bits for quantization.
Further, we use the same codebooks for each individual scheme which are designed by the \textit{Lloyd algorithm} \cite[Chapter 6]{91:Gersho}. Further, we initialize \algref{alg1} with the same codebook sets for the nearest-neighbor coding. Using the support set coding, a codebook set is designed for a Gaussian scalar component. Hereafter, we make a convention that $\textrm{NMSE} = 1$ for the support set coding scheme if the total available bit-budget is $R_x < K \log_2 M$ bits.

\begin{figure}
  \begin{center}
  \includegraphics[width=\columnwidth,height=5.55cm]{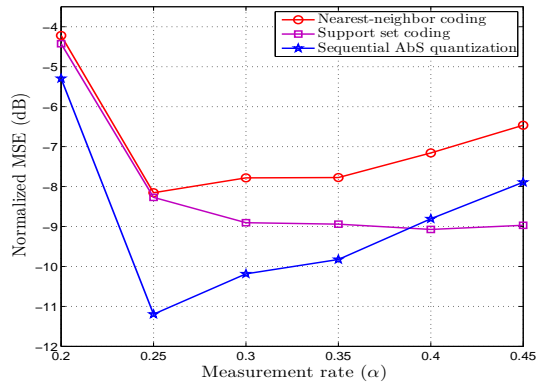}\\
  \vspace{-0.5cm}
  \caption{NMSE as a function of measurement rate ($\alpha = N/M$).}
  \label{fig:result2}
  \end{center}
  \vspace{-0.85cm}
\end{figure}

The performance of the qunatization algorithms as a function of measurement rate $\alpha$ is reported in \figref{fig:result2}. Let us first consider the nearest-neighbor coding and the proposed AbS-based quantization scheme. From the curves, it can be observed that given a very small values of $\alpha$, the OMP algorithm fails to detect the sparsity pattern and reconstruct the source which results in a poor performance although the quantization rate per entry is high. As $\alpha$ increases to a certain amount, the reconstruction algorithm succeeds to reconstruct the sparse source precisely out of the measurements since the number of measurements is sufficient, and the quantization error is small enough. At this point ($\alpha = 0.25$ for \figref{fig:result2}), the curves reach the best performance. However, for higher $\alpha$'s, due to the limited quantization rate $r_y$, the quantization error per entry increases which leads to a poorer performance. Next, we evaluate the performance of the support set coding where the quantization is performed on the locally reconstructed signal domain. Similarly, at small $\alpha$'s, the OMP reconstruction algorithm fails to reconstruct the locally sparse source, where the performance is insignificant. It can be seen that as $\alpha$ increases and the OMP is able to reconstruct the input signal vector, the performance improves slightly by further increasing measurement rate since the allocated quantization bits using this method are independent of number of measurements. In the spirit of exploiting CS for practical applications, we are mainly interested in the lower ranges of $\alpha$ (e.g. $\alpha = 0.25$), where the AbS-based quantization scheme achieves a considerable $3$ dB reduction in MSE.

\begin{figure}
  \begin{center}
  \includegraphics[width=\columnwidth,height=5.55cm]{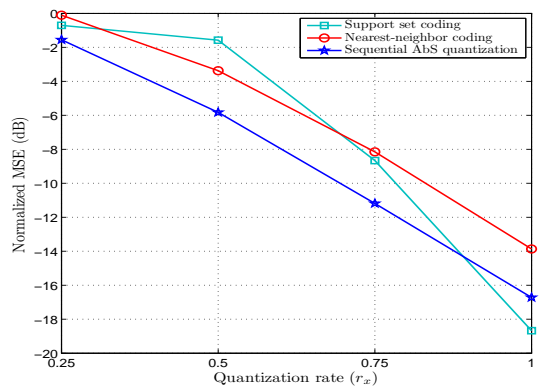}\\
  \vspace{-0.5cm}
  \caption{NMSE as a function of quantization rate $r_x$.}
  \label{fig:result4}
  \end{center}
  \vspace{-1cm}
\end{figure}

Now, we show the performance (NMSE) as a function of quantization rate per entry of $\mathbf{X}$ (i.e., $r_x$) in \figref{fig:result4} using parameters $M=512$, $K=35$ at $\alpha=0.25$. It can be observed that at low to moderate-high rates, the AbS quantization outperforms the other schemes, while at high rates the support set coding attains a slightly better performance. At very high rates, the performance of the support set coding is expected to saturate since, at $\alpha=0.25$, the distortion due to recovery of the locally reconstructed vector $\tilde{\mathbf{x}}$, according to which the quantization is performed, remains constant. In total, the AbS-based quantization outperforms at operational ranges of measurement and quantization rates where the available resources (number of sensors and quantization rate) are constrained.

\vspace{-0.25cm}
%%%%%%%%%%%%%%%%%%%%%%%%%%%%%%%%%%%%%%%%%%%%%%%%%%%%%%%%%%%%%%%%%%%%%%%%%%%%%%%%%%%%%
\section{Conclusions} \label{sec:conclusion}
%%%%%%%%%%%%%%%%%%%%%%%%%%%%%%%%%%%%%%%%%%%%%%%%%%%%%%%%%%%%%%%%%%%%%%%%%%%%%%%%%%%%%
\vspace{-0.25cm}
We have developed a new framework of quantizer design for CS measurements. We have considered a resource-constrained application where both measurements and transmission rates are limited. Using this scenario, we have addressed the problem of encoding CS measurements where inspired by the AbS framework, a new quantization algorithm has been proposed for coding of linear CS measurements. Numerical results have shown the promising performance gain obtained using this scheme.

\bibliographystyle{IEEEtran}
\bibliography{IEEEfull,bibliokthPasha}
\end{document}